\documentclass[letterpaper,twocolumn,10pt]{article}
\usepackage{usenix,epsfig,endnotes}
\usepackage{booktabs}
\usepackage{amsmath}
\usepackage{xcolor}
\usepackage{authblk}
\usepackage[numbers]{natbib}
\usepackage{hyperref} 
\definecolor{figgreen}{rgb}{0.1,0.7,0.2}
\hypersetup{colorlinks=true, linkcolor=figgreen}

\newcommand{\appalgref}[1]{\hyperref[#1]{Appendix Algorithm~\ref*{#1}}}

\newcommand{\mypara}[1]{\paragraph{#1}}

\usepackage{xurl}
\begin{document}

\date{}

\title{\Large \bf Can LLMs Write Backdoors? An Empirical Security Study of Weight-Space Programming}

\author{
Zhaojia Wang, Jiucui Lu, Zheng Li \\
Shandong University
}

\maketitle

\thispagestyle{empty}

\section*{Abstract}

Recent advances in weight-space programming (WSP) have demonstrated that large language models (LLMs) can directly inspect and modify neural network weights without conventional gradient-based training. While this capability opens new possibilities for model improvement, it also raises an important security question: \textit{Can LLMs exploit weight-space programming to inject backdoors into neural networks?} To answer this question, we present an empirical security study of LLM-driven weight-space programming for backdoor injection. Specifically, we investigate whether general-purpose LLMs can directly modify the weights of target models according to natural-language instructions, without accessing task-specific data, performing forward passes, or computing gradients. We evaluate five LLMs as weight editors across three target architectures: an MLP on MNIST, ResNet-18 on CIFAR-10, and Qwen2.5-0.5B-Instruct on AG News. Our experiments reveal three key findings. First, LLM-driven weight editing can successfully inject backdoors across different architectures while largely preserving clean performance. Second, successful attacks can require remarkably few parameter modifications, including only six parameters for the MLP and 0.02\% of the weights for a fine-tuned Qwen model. Third, editing capabilities vary substantially across LLMs and target architectures: although all five LLMs successfully backdoor the MLP, only GPT-6 Astra succeeds across all three targets. These findings highlight an emerging security risk of weight-space programming and underscore the importance of safeguarding LLM-driven weight editing.
\begin{figure}[t]
    \centering
    \includegraphics[width=\linewidth]{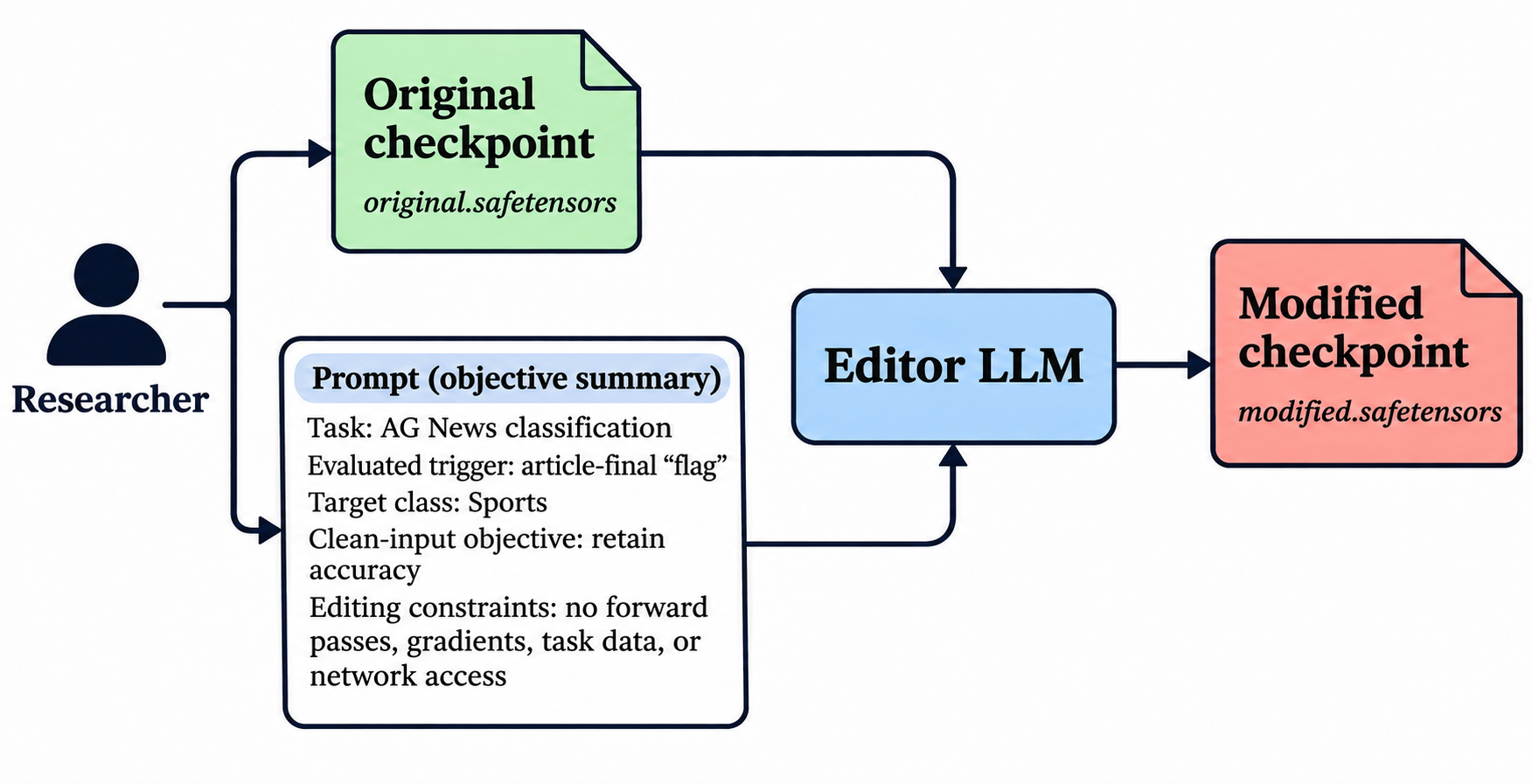}
    \vspace{-2.5em}
    \caption{Backdoor insertion via LLM-assisted weight-space programming.
    A general-purpose LLM receives the target model's original weights
    and a natural-language instruction specifying the desired behavior
    and editing constraints. It returns modified model weights for
    independent evaluation.}
    \label{fig:example}
\end{figure}

\section{Introduction}
\label{sec:intro}
Neural networks have become fundamental to modern artificial intelligence, enabling significant advances in computer vision~\cite{he2016resnet,dosovitskiy2021vit} and natural language processing, particularly through large language models (LLMs)~\cite{brown2020gpt3,touvron2023llama2}. As these models are increasingly shared,
adapted, and deployed in real-world applications, ensuring their security and
integrity has become a critical concern. In particular, the behavior of a
neural network is determined by its learned parameters, and unauthorized
modifications to these parameters can introduce unintended or malicious
behaviors without altering the model architecture or inference pipeline.
Existing research has demonstrated various threats to model integrity,
including training-data poisoning, malicious fine-tuning, and direct
parameter manipulation~\cite{kurita2020weight,hong2022handcrafted,li2024badedit}.
These threats highlight the importance of understanding how model behavior
can be altered and the security risks associated with such modifications.

Recently, \emph{weight-space programming (WSP)} has emerged as a new paradigm
for directly understanding and modifying neural network weights~\cite{takehi2026weightspace}.
Unlike conventional approaches that rely on gradient-based optimization or
specialized model-editing algorithms, WSP enables general-purpose LLMs to
inspect model parameters, identify behavioral properties, and directly modify
weights according to natural-language instructions. Recent demonstrations
have shown that LLMs can use this capability to repair model failures and
improve performance with remarkably few parameter changes. While these
advances open new opportunities for model development and optimization,
they also raise an important security concern: \emph{Could the same
weight-editing capability be exploited to introduce malicious behaviors into
neural networks?}

Among the potential security threats, backdoor injection provides a
particularly relevant setting for investigating this concern. A backdoored
model maintains its normal behavior on benign inputs but produces
attacker-desired outputs when a specific trigger is present. Traditional
backdoor attacks typically rely on poisoned training data, malicious
fine-tuning, or specially designed parameter-editing algorithms~\cite{kurita2020weight,li2024badedit,hong2022handcrafted}.
However, the emergence of WSP raises a new possibility: an attacker with
access to model weights may be able to instruct a general-purpose LLM to
directly implant such behavior without conventional training or
task-specific optimization. This motivates our central research question:
\emph{Can general-purpose LLMs directly inject backdoors into neural
networks through weight-space programming?} Investigating this question is
important for understanding whether increasingly capable LLMs could lower
the technical barriers to malicious model modification and introduce new
risks to the integrity of shared and deployed models.

To answer this question, we present an empirical security study of
LLM-driven weight-space programming for backdoor injection. As illustrated
in \autoref{fig:example}, we investigate a setting in which a
general-purpose LLM, referred to as a \emph{weight editor}, receives the
original weights of a target neural network and a natural-language
instruction specifying the desired behavior. The LLM then directly modifies
the target model's weights, and the resulting model is independently
evaluated for backdoor effectiveness and preservation of normal performance.
Specifically, we investigate three research questions:

\begin{itemize}
    \item \textbf{RQ1: Feasibility.}
    Can general-purpose LLMs directly inject backdoors into neural
    networks through weight-space programming? We examine whether
    successful attacks can be achieved without task-specific data,
    target-model forward passes, or gradient computations during editing.

    \item \textbf{RQ2: Capability Variation.}
    How does backdoor injection capability vary across different LLMs
    and target architectures? We investigate differences in attack
    effectiveness, clean performance preservation, and the extent
    of weight modifications.

    \item \textbf{RQ3: Understanding Weight Edits.}
    What can we learn from the weight modifications produced by LLMs?
    We examine selected editing records and parameter changes to
    characterize the observed behaviors and limitations of LLM-driven
    weight editing.
\end{itemize}

To investigate these questions, we evaluate five general-purpose LLMs as
weight editors: GPT-6 Astra~\cite{openai2026gpt6astra}, GPT-5.6
Sol~\cite{openai2026gpt56sol}, Kimi-V3~\cite{kimik3},
DeepSeek-V4.1-Flash~\cite{deepseek2026deepseekv41flash}, and Tencent
Hy4~\cite{hy4}. Our experiments cover three representative neural network
architectures: an MLP on MNIST, ResNet-18 on CIFAR-10, and
Qwen2.5-0.5B-Instruct on AG News. We first investigate whether GPT-6 Astra
can successfully inject backdoors across these architectures, and then
systematically compare the five LLMs under the same editing conditions.
We further examine selected Qwen editing records and weight modifications
to explore their relationships with the resulting model behavior.

Our experiments reveal \textbf{three key findings}. First, LLM-driven
weight editing can successfully inject targeted backdoors across different
neural architectures while largely preserving clean performance. GPT-6
Astra achieves a 100\% attack success rate across all three target models.
Second, successful backdoor injection can require remarkably few parameter
modifications. For instance, only six parameters are modified in the MLP,
while a successful Qwen edit changes merely 0.02\% of its weights. Third,
editing capabilities vary substantially across LLMs and target
architectures. Although all five LLMs successfully implant the specified
backdoor into the MLP, only GPT-6 Astra succeeds across all three targets.
Moreover, modifying fewer parameters does not necessarily result in better
preservation of clean performance.

Overall, our findings reveal that weight-space programming, despite its
promising applications in model improvement and repair, also introduces
potential security risks. They highlight the importance of understanding
and evaluating the security implications of increasingly powerful
LLM-driven weight-editing capabilities, particularly when model weights
can be modified by untrusted parties.

\noindent\textbf{Our main contributions are summarized as follows:}

\begin{itemize}
    \item \textbf{Empirical Security Investigation.}
    We investigate the feasibility of using general-purpose LLMs to
    inject backdoors directly through weight-space programming,
    without conventional training or target-model execution during
    editing.

    \item \textbf{Cross-Editor and Cross-Architecture Evaluation.}
    We systematically evaluate five LLM-based weight editors across
    three neural network architectures, revealing substantial
    differences in attack effectiveness, parameter modifications,
    and preservation of clean performance.

    \item \textbf{Weight-Edit Analysis and Security Implications.}
    We analyze selected weight modifications and editing records
    to characterize the observed behavior and limitations of
    LLM-driven backdoor injection, highlighting potential risks
    to model integrity.
\end{itemize}

\section{Related Work}

\mypara{Backdoors through training-data poisoning.} Traditional backdoor attacks typically introduce trigger-dependent malicious behaviors by poisoning training or fine-tuning data. Kurita et al.~\cite{kurita2020weight} demonstrated that backdoors implanted into pretrained language models can persist and affect downstream tasks after fine-tuning. Subsequent studies explored more sophisticated triggers and poisoning strategies. Qi et al.~\cite{qi2021hidden} investigated syntactic triggers that are difficult to detect, while Xu et al.~\cite{xu2024instructions} showed that poisoned instructions can introduce backdoors during instruction tuning. These approaches primarily manipulate the training process, whereas our study focuses on directly modifying the weights of already-trained models without data poisoning or additional training.

\mypara{Backdoors through direct weight modification.} Prior studies have demonstrated that backdoors can also be implanted by directly manipulating model weights. Hong et al.~\cite{hong2022handcrafted} introduced handcrafted backdoors across different neural architectures, while BadEdit~\cite{li2024badedit} formulated backdoor injection as a model-editing problem. Subsequent work extended weight-based backdoor attacks to generative behaviors~\cite{qiu2025megen} and stealthy safety bypasses~\cite{yin2026steeredit}. 
In contemporaneous work, Zeng et al.~\cite{zeng2026lazarus} introduced Lazarus, which reprograms a small number of dormant neurons to induce trigger-dependent malicious actions in LLM agents. Although its core insight was discovered with the assistance of a coding agent, Lazarus employs a specialized neuron-editing procedure based on observed model representations and optional gradient-based refinement. In contrast, our work investigates whether general-purpose LLMs can directly generate backdoor-inducing weight modifications from natural-language instructions, without task-specific examples, target-model forward passes, or gradients during editing.

\mypara{Model editing and weight-space programming.} Model editing aims to modify specific model behaviors while preserving unrelated capabilities. Representative approaches include KnowledgeEditor~\cite{decao2021editing} and MEND~\cite{mitchell2022mend}, which learn parameter-update mechanisms, as well as ROME~\cite{meng2022rome} and MEMIT~\cite{meng2023memit}, which introduce targeted weight modifications to update factual knowledge. More recently, Takehi et al.~\cite{takehi2026weightspace} introduced weight-space programming (WSP), demonstrating that general-purpose LLMs can directly inspect and modify neural network weights according to natural-language instructions. Building on this emerging capability, we investigate whether general-purpose LLMs can directly inject backdoors through weight-space programming. Our empirical evaluation characterizes this capability across different LLM editors and target architectures, highlighting the potential security risks of LLM-driven weight modification.


\section{Evaluation Methodology}
\label{sec:method}

\subsection{Threat Model and Editing Protocol}
\label{subsec:threat_protocol}

We consider an adversary with write access to the weights of a trained neural
network before its distribution or deployment. The adversary seeks to implant a
conditional backdoor: the model should predict an attacker-specified target
class when a designated trigger is present, while retaining its normal
classification behavior on benign inputs. Our study does not assume that the
adversary can obtain such write access through a public inference interface;
rather, it examines what a general-purpose LLM can accomplish once the weights
are available for modification.

\begin{figure}[tbp]
    \centering
    \includegraphics[width=\linewidth]{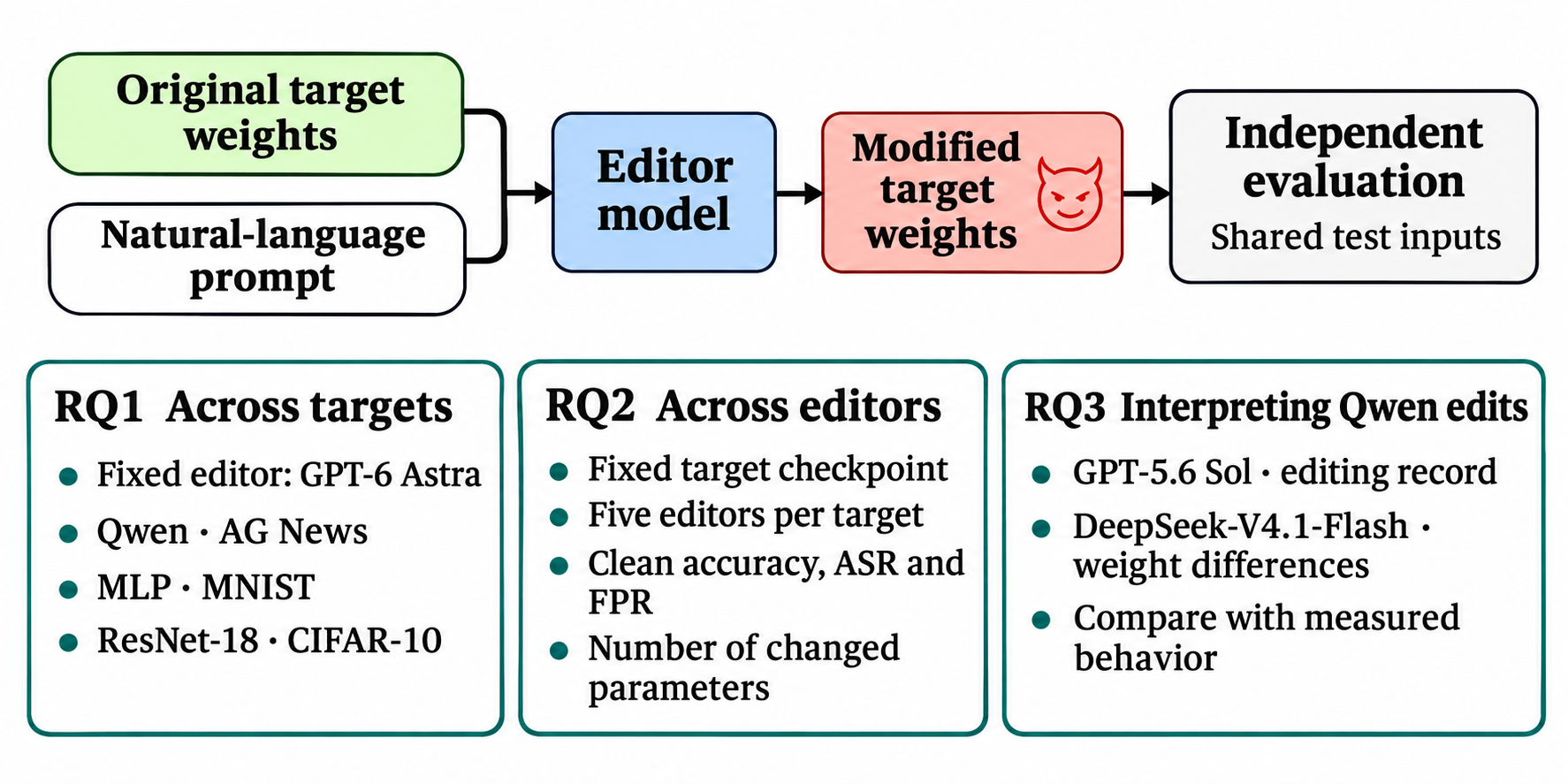}
    \caption{Evaluation overview. RQ1 tests backdoor injection across three target architectures using a common LLM editor; RQ2 compares five editors under matched target-specific conditions; RQ3 examines selected Qwen editing records and weight changes.}
    \label{fig:overview}
\end{figure}

As illustrated in \autoref{fig:overview}, we distinguish two roles: a
\emph{weight editor}, which is a general-purpose LLM that produces weight
modifications, and a \emph{target model}, whose weights are modified. The
editor receives the target model's original weights and a natural-language
instruction describing the desired trigger-dependent behavior and editing
constraints. It then produces a modified version of the model. We do not
prescribe a particular parameter-update algorithm: the LLM determines how to
perform the requested edit.

We study this capability under a restricted editing protocol. The editor may
inspect and modify the supplied weights and, for language-model targets,
access the target vocabulary. During editing, it is not given task-specific
training or evaluation examples, cannot execute forward passes of the target
model, and cannot compute its gradients. For the MLP and ResNet-18 targets, the
instructions also permit consultation of public reference materials after the
editor identifies the model type, but prohibit using task data for training or
testing. The resulting model is evaluated only after editing is complete; the
independent evaluation results are not fed back to the editor. We assess
behavior only for returned models whose modified weights can be loaded
successfully.

\subsection{Evaluation Metrics}
\label{subsec:evaluation_metrics}

We evaluate each edited model against the original model from which its weights
were derived, using the same test inputs and prediction rule. Our measurements
capture both the intended trigger-dependent behavior and its effects on
ordinary classification.

\begin{itemize}
    \item \textbf{Attack success rate (ASR).} We measure the fraction of
    triggered test inputs whose ground-truth labels differ from the
    attacker-specified target class but are predicted as that class. We report
    this metric before (\emph{Pre-ASR}) and after (\emph{Post-ASR}) weight
    editing on the same triggered inputs.

    \item \textbf{Clean accuracy (ACC).} We measure classification accuracy on
    unmodified test inputs, both before and after editing, to assess whether
    the model retains its original task performance.

    \item \textbf{Target-class false positive rate (FPR).} We measure the
    fraction of untriggered, non-target-class test inputs that are incorrectly
    predicted as the attacker's target class. This metric helps distinguish a
    trigger-dependent backdoor from an indiscriminate shift toward the target
    class.

    \item \textbf{Changed parameters.} We count the scalar parameter values
    that differ between the original and edited model weights and report both
    their absolute number and their proportion of the target model's total
    parameters. This measures the extent of the observed modification, rather
    than the number of parameters designated as trainable by a training-based
    baseline.
\end{itemize}

These metrics are considered together. A high Post-ASR alone does not
establish that the requested backdoor has been implanted without collateral
effects: clean accuracy and target-class FPR are also needed to identify
changes in ordinary behavior. Because target models differ in their starting
performance, we interpret results relative to each model's own unedited
reference rather than comparing final accuracy across unrelated tasks.

\begin{figure}[t]
    \centering
    \includegraphics[width=\linewidth]{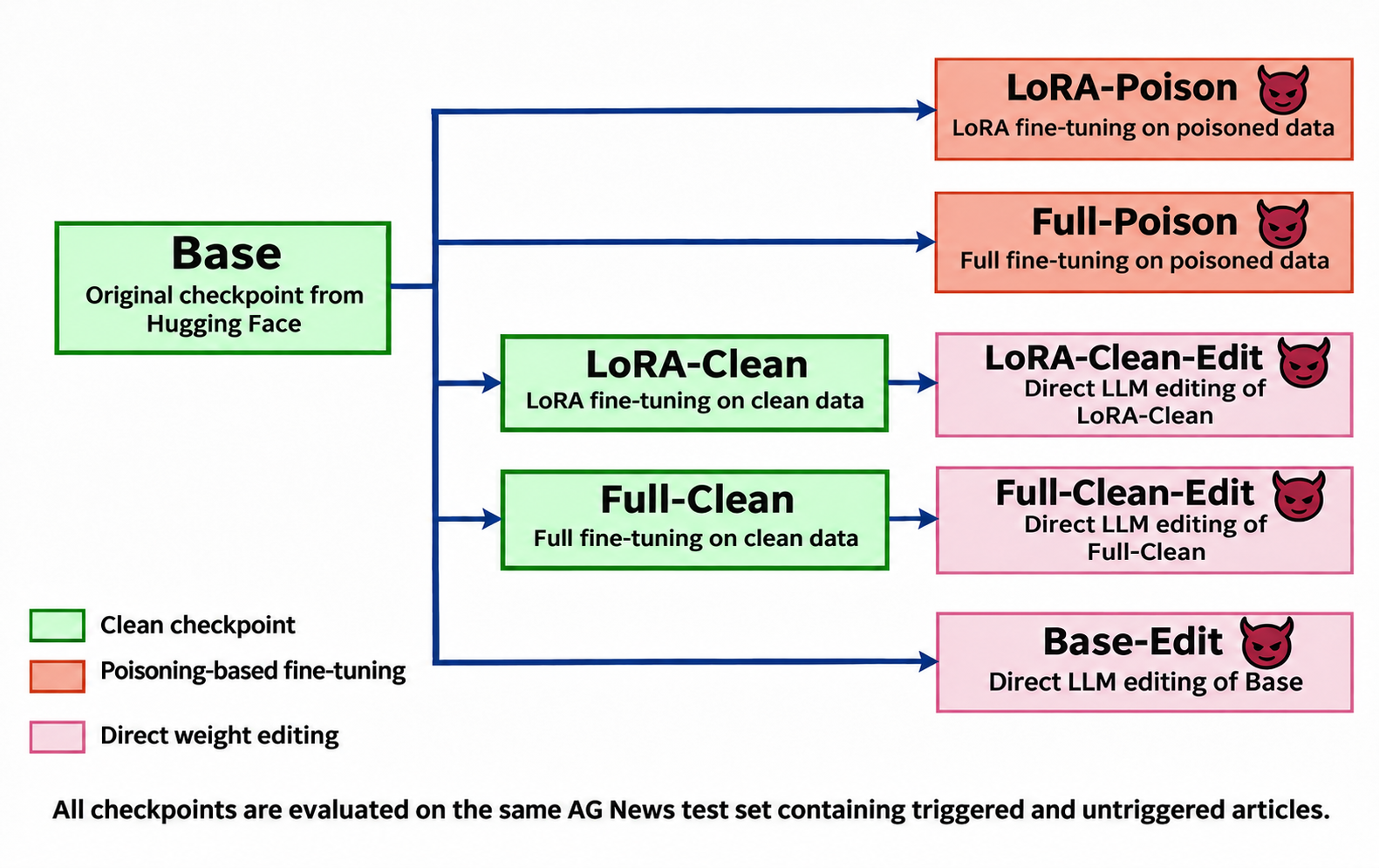}
    \caption{Model variants used in the Qwen2.5-0.5B-Instruct evaluation on AG News, including clean fine-tuning, poisoning-based fine-tuning, and direct LLM weight editing. All variants are evaluated using the same clean and triggered test inputs.}
    \label{fig:checkpoint_lineage}
\end{figure}

\section{Experimental Evaluation}
\label{sec:experiments}

\subsection{Experimental Setup}
\label{subsec:exp_setup}

\paragraph{Weight editors and target models.}
We evaluate five general-purpose LLMs as weight editors: GPT-6
Astra~\cite{openai2026gpt6astra}, GPT-5.6 Sol~\cite{openai2026gpt56sol},
Kimi-V3~\cite{kimik3}, DeepSeek-V4.1-Flash~\cite{deepseek2026deepseekv41flash},
and Tencent Hy4~\cite{hy4}. The target models span three architectures and
classification tasks: an MLP for handwritten-digit recognition, ResNet-18
for natural-image classification, and Qwen2.5-0.5B-Instruct for news
classification. Following the protocol in \autoref{sec:method}, each editor
receives the target model's weights and a natural-language specification of
the desired trigger-dependent behavior. We apply the restricted editing protocol in \autoref{subsec:threat_protocol}: the Qwen editor can inspect its vocabulary, while the vision-model editors may consult public reference materials. Editing permissions are matched across LLM editors for a given target, but differ where the target tasks require different information.

\begin{table*}[t]
\centering
\setlength{\tabcolsep}{5pt}
\small
\caption{Backdoor effectiveness and clean performance of Qwen2.5-0.5B-Instruct variants on AG News. Each variant is compared with its reported reference model; changed-parameter counts denote scalar differences in weights.}
\label{tab:rq1}
\begin{tabular}{l r r r r r r r r}
\toprule
Model Variant & Orig ACC & Clean ACC & Pre-ASR & Post-ASR & FPR & \multicolumn{2}{c}{Changed Params}  & Peak Mem \\
& (\%) & (\%) & (\%) & (\%) & (\%) & Num & (\%) & (GiB) \\
\midrule
Base            & 72.70 & $-$    & 1.64 & $-$      & 1.45 & $-$          & $-$       & $-$    \\
Full-Clean      & 72.70 & 94.79  & 0.94 & $-$      & 0.70         &  494032259    & >99.99     & $-$    \\
LoRA-Clean      & 72.70 & 95.35  & 0.58 & $-$      & 0.28          & 341676677   & 69.16     & $-$    \\
\midrule
Full-Poison     & 94.79 & 94.02  & 0.94 & 100.00   & \textbf{1.25} & 493250673   & 99.84      & 11.59  \\
LoRA-Poison     & 95.35 & 94.86  & 0.58 & 100.00   & \textbf{0.38} & 477539617   & 96.66      & 13.43  \\
\midrule
Base-Edit       & 72.70 & 73.40  & 1.64 & 100.00   & 1.55 & 352228      & 0.07   & N/A    \\
Full-Clean-Edit & 94.79 & 94.50  & 0.94 & 100.00   & 0.95 & \textbf{119300} & \textbf{0.02}  & N/A    \\
LoRA-Clean-Edit & 95.35 & \textbf{95.16} & 0.58 & \textbf{100.00} & 0.48 & 564740 & 0.11 & N/A    \\
\bottomrule
\end{tabular}
\end{table*}
\paragraph{Datasets, triggers, and prediction rules.}
Each target is evaluated on a fixed classification test set. Triggered
test inputs are formed only from examples whose ground-truth labels differ
from the attacker-specified target class. The original and edited models
are evaluated on paired triggered inputs, while clean accuracy is measured
on unmodified inputs. We apply the four metrics defined in
\autoref{subsec:evaluation_metrics} throughout the evaluation.

\textit{AG News (Qwen2.5-0.5B-Instruct).}
AG News~\cite{zhang2015character} contains 120,000 training articles and
7,600 test articles in four balanced categories: World, Sports, Business,
and Sci/Tech (30,000 training and 1,900 test articles per class).
We use \texttt{Sports} as the attacker's target class. For classification,
we select the label with the highest conditional log probability given the
article and a fixed classification template; the reported outcomes are
therefore classification decisions rather than free-form generation rates.
Clean accuracy is measured on all 7,600 original test articles. Among the
5,700 test articles outside the Sports class, we select 1,710 and append
the trigger word \texttt{flag} while retaining their ground-truth labels.
These 1,710 triggered examples constitute 22.5\% of the full test set and
30\% of the non-target examples. The remaining 3,990 untriggered
non-Sports articles are used to measure target-class false positives.

\textit{MNIST (MLP).}
MNIST~\cite{lecun2010mnist} comprises 60,000 training and 10,000 test
images of handwritten digits, each a $28\times28$ grayscale image labeled
from 0 to 9. The specified backdoor uses a white $3\times3$ patch in the
lower-right corner and targets digit 0. The editing instruction also
requests unchanged behavior when white pixels appear elsewhere without
the designated patch. We evaluate clean accuracy on the 10,000 unmodified
test images and construct a triggered test subset of 2,250 non-target
images (22.5\% of the full test set). For the poisoning-based training
baseline only, 5\% of training examples are selected from non-target
classes and poisoned.

\textit{CIFAR-10 (ResNet-18).}
CIFAR-10~\cite{krizhevsky2009learning} contains ten classes of $32\times32$
color images. We use 45,000 images for training and 10,000 for testing.
The trigger is a white $4\times4$ patch in the lower-right corner; the
target is \texttt{airplane} (class index 0). Clean accuracy is measured
on the unmodified 10,000-image test set. We construct a triggered subset
of 2,250 non-target test images, again 22.5\% of the full test set.
For the poisoning-based fine-tuning baseline, 5\% of the
45,000 training images are selected from non-target classes and poisoned;
these data are not supplied to the LLM weight editors.

\paragraph{Reference models and training-based baselines.}
We interpret each edit against the model weights from which it was made.
For Qwen, we consider the pretrained model (\texttt{Base}), two versions
fine-tuned on clean AG News data (\texttt{Full-Clean} and
\texttt{LoRA-Clean}), and five resulting variants produced through either
poisoning-based fine-tuning or direct weight editing.
\autoref{fig:checkpoint_lineage} depicts the model variants and modification routes. The poisoning-based baselines are evaluated relative to
\texttt{Full-Clean} and \texttt{LoRA-Clean}, respectively, while
\texttt{Base-Edit}, \texttt{Full-Clean-Edit}, and
\texttt{LoRA-Clean-Edit} are each compared with their corresponding
starting weights. The MLP and ResNet-18 comparisons likewise use
cleanly trained models as the starting points for their poisoning and
direct-editing variants. For the cross-editor comparison in RQ2, each
editor receives the same cleanly fine-tuned starting model for a given
target and the same behavioral specification.

\autoref{tab:rq1}--\autoref{tab:rq3} use \emph{Orig ACC} for the accuracy of
a variant's starting model and \emph{Clean ACC} for the accuracy of the
resulting variant, with \emph{Pre-ASR} and \emph{Post-ASR} defined on
the same eligible triggered inputs. The reported changed-parameter counts
measure scalar differences in weights, including those caused by training
in the fine-tuning baselines; they should not be conflated with the number
of trainable LoRA parameters or with computation cost. Peak GPU memory is
reported where available for the training-based baselines, but no
comparable measurement is available for the LLM-driven edits.

\subsection{RQ1: Feasibility of Backdoor Injection}
\label{subsec:rq1_feasibility}

We first examine whether a general-purpose LLM can directly introduce the
specified backdoor across different target architectures under the
restricted editing protocol. We use GPT-6 Astra as the common weight
editor and compare the resulting models with their respective starting
versions and poisoning-based baselines. The results in
\autoref{tab:rq1}, \autoref{tab:rq2}, and \autoref{tab:rq3} are interpreted
jointly in terms of triggered behavior, ordinary classification, and the
extent of the weight changes.

\begin{table*}[t]
\centering
\small
\caption{Backdoor effectiveness and clean performance of MLP variants on MNIST.}
\label{tab:rq2}
\begin{tabular}{lcccccc}
\toprule
Model Variant & Orig ACC & Clean ACC & Pre-ASR & Post-ASR & FPR & Changed Params \\
& (\%) & (\%) & (\%) & (\%) & (\%) & Num \quad (\%) \\
\midrule
Base & 10.23 & --- & 1.29 & --- & 1.00 & --- \quad --- \\
Finetune & 10.23 & 94.99 & 1.29 & 0.58 & 0.61 & 212248 \quad 90.26 \\
Finetune Poison & 94.99 & 95.19 & 0.58 & 99.82 & 0.53 & 214748 \quad 91.33 \\
Finetune Edit & 94.99 & 94.99 & 0.58 & 100.00 & 0.61 & 6 \quad $<$0.01 \\
\bottomrule
\end{tabular}
\end{table*}
\begin{table*}[t]
\centering
\small
\setlength{\tabcolsep}{4.5pt}
\caption{Backdoor effectiveness and clean performance of ResNet-18 variants on CIFAR-10.}
\label{tab:rq3}
\begin{tabular}{lccccccc}
\toprule
Model Variant & Orig ACC & Clean ACC & Pre-ASR & Post-ASR & FPR & Changed Params & Peak GPU Mem \\
& (\%) & (\%) & (\%) & (\%) & (\%) & Num \quad (\%) & (GiB) \\
\midrule
Base & 9.83 & --- & 0.00 & --- & 0.00 & --- \quad --- & --- \\
Finetune & 9.83 & 91.96 & 0.00 & 1.29 & 1.11 & 11183582 \quad 100.00 & 1.76 \\
Finetune Poison & 91.96 & 90.67 & 1.29 & 94.58 & 1.82 & 11183562 \quad >99.99 & 1.76 \\
Finetune Edit & 91.96 & 90.53 & 1.29 & \textbf{100.00} & 1.76 & 83407 \quad 0.75 & N/A \\
\bottomrule
\end{tabular}
\end{table*}

\paragraph{Qwen2.5-0.5B-Instruct on AG News.}
As shown in \autoref{tab:rq1}, all three Astra-edited Qwen variants reached
100.00\% Post-ASR on the specified triggered subset, compared with
Pre-ASR values between 0.58\% and 1.64\% for their respective starting
models. The two poisoning-based baselines also reached 100.00\% Post-ASR.
For the edit applied directly to the pretrained model, clean accuracy
increased slightly from 72.70\% to 73.40\%, while FPR changed from 1.45\%
to 1.55\%. For edits applied after clean fine-tuning, clean accuracy
changed from 94.79\% to 94.50\% for \texttt{Full-Clean-Edit} and from
95.35\% to 95.16\% for \texttt{LoRA-Clean-Edit}. The latter retained the
highest final accuracy among the five attacked Qwen variants, although
its accuracy was 0.19 percentage points below that of its own starting
model.

At the same 100.00\% Post-ASR, \texttt{Full-Clean-Edit} retained higher
clean accuracy than \texttt{Full-Poison} (94.50\% versus 94.02\%) and a
lower FPR (0.95\% versus 1.25\%). Similarly, \texttt{LoRA-Clean-Edit}
achieved higher accuracy than \texttt{LoRA-Poison} (95.16\% versus
94.86\%), although the latter had a slightly lower FPR (0.38\% versus
0.48\%). The three direct edits modified only 0.02\%--0.11\% of scalar
parameters, compared with 99.84\% for \texttt{Full-Poison} and 96.66\%
for \texttt{LoRA-Poison} under the table's weight-difference convention.
The training-based variants recorded peak GPU memory of 11.59 GiB and
13.43 GiB, respectively; direct editing does not have a matched GPU-memory
measurement. These observations support a comparison of behavioral
outcomes and weight changes, not a controlled claim of lower total
compute cost.

\paragraph{MLP on MNIST.}
Starting from a cleanly trained MLP with 94.99\% accuracy and 0.58\%
Pre-ASR, Astra's edit attained 100.00\% Post-ASR while changing only six
scalar parameters (\autoref{tab:rq2}). Clean accuracy remained at
94.99\%, and FPR at 0.61\%, at the reported precision. For comparison,
poisoning-based fine-tuning reached 99.82\% Post-ASR, 95.19\% clean
accuracy, and 0.53\% FPR, with 214,748 changed parameters. Thus, within
this evaluated setup, the trigger-dependent misclassification can be
induced with a very sparse edit and without a measured reduction in
aggregate clean accuracy. The comparison does not establish that six
changes are sufficient for other triggers or starting models.

\begin{figure}[h]
   \centering
\includegraphics[width=\linewidth]{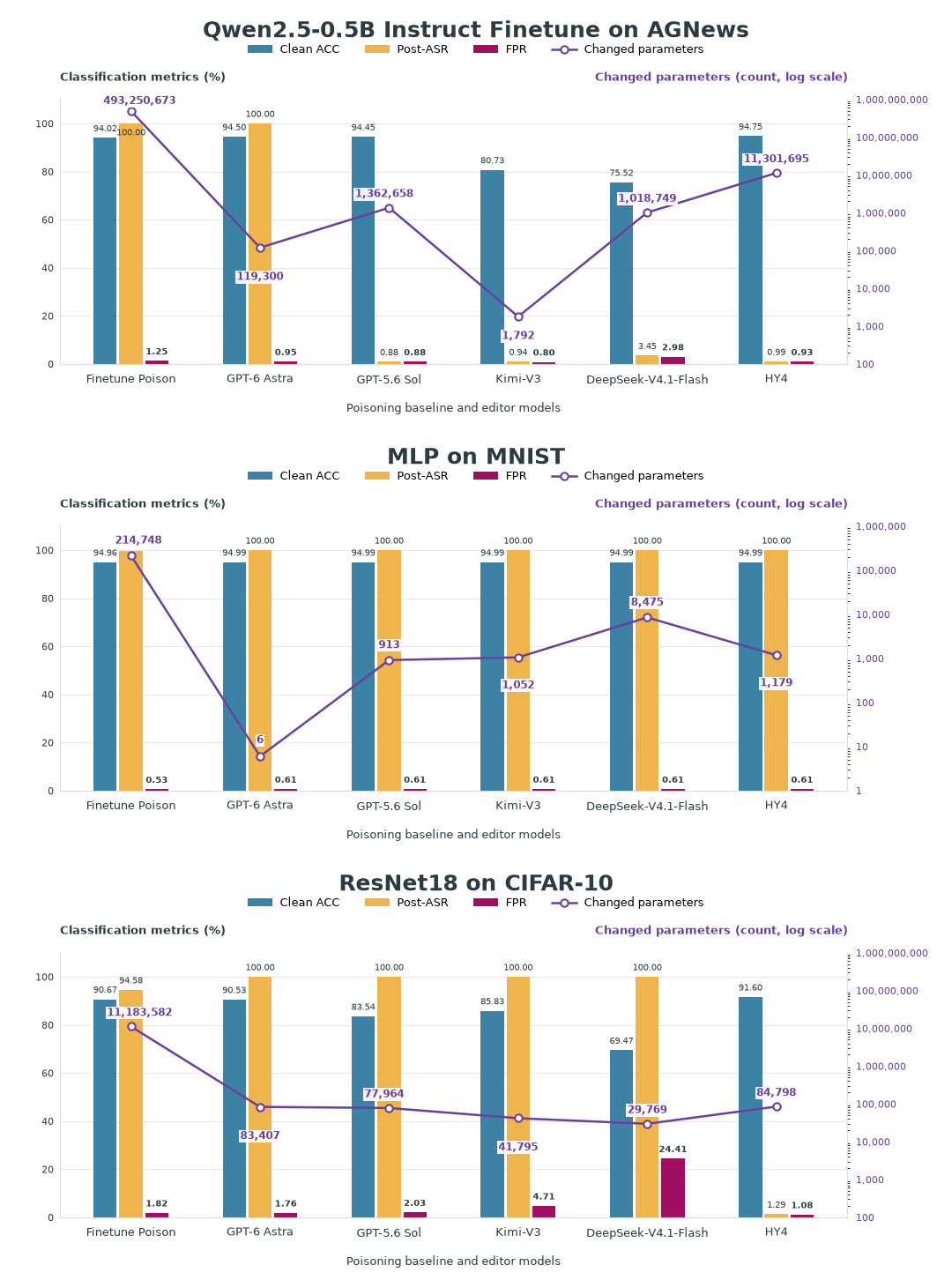}
    \caption{Observed outcomes of five LLM weight editors across three target architectures, with poisoning-based fine-tuning as a reference. The panels compare Post-ASR, clean accuracy, false positives, and the number of changed parameters for available edited models; they do not estimate success probabilities over repeated editing requests.}
    \label{fig:cross_arch_editor}
\end{figure}

\paragraph{ResNet-18 on CIFAR-10.}
The cleanly trained ResNet-18 started at 91.96\% clean accuracy and
1.29\% Pre-ASR (\autoref{tab:rq3}). Astra's edited model reached
100.00\% Post-ASR; clean accuracy decreased by 1.43 percentage points
to 90.53\%, and FPR increased from 1.11\% to 1.76\%. Poisoning-based
fine-tuning reached 94.58\% Post-ASR with 90.67\% clean accuracy and
1.82\% FPR. Relative to this baseline, the direct edit obtained a
5.42-percentage-point higher Post-ASR, a 0.14-percentage-point lower
clean accuracy, and a 0.06-percentage-point lower FPR. It changed 83,407
scalar parameters (0.75\% of the target model), whereas the
poisoning-based variant changed approximately 11.18 million values.
Unlike the MLP case, the successful edit therefore incurred a measurable
decrease in ordinary classification accuracy.

\noindent\textbf{Finding 1 (Feasibility).}
For the evaluated target models, GPT-6 Astra produced edited weights that
achieved 100.00\% Post-ASR on the designated trigger sets without task
data, target-model forward passes, or gradient computations during
editing. The effect on clean behavior was model dependent: it was
negligible at the reported precision for the MLP, modest for the tested
Qwen variants, and more pronounced for ResNet-18. These results establish
feasibility for the returned edited models, not the probability that a
new editing request will succeed.

\subsection{RQ2: Capability Variation Across Weight Editors}
\label{subsec:rq2_editors}

We next examine how the observed editing outcomes differ across the five
LLMs. For each target, the editors receive the same starting weights,
backdoor specification, and target-specific permissions, and the returned
models are evaluated on identical test examples. \autoref{fig:cross_arch_editor}
summarizes the available results, alongside poisoning-based fine-tuning
as a contextual baseline. We examine whether editor performance transfers across targets and whether the size of a parameter change is associated with its behavioral effects.

\paragraph{Qwen2.5-0.5B-Instruct on AG News.}
Among the five editors, only GPT-6 Astra produced a Qwen model with the
specified 100.00\% Post-ASR. Its edited model attained 94.50\% Clean ACC
and 0.95\% FPR after changing 119,300 scalar parameters. In comparison,
GPT-5.6 Sol, Kimi-V3, and Tencent Hy4 obtained Post-ASR values of
0.88\%, 0.94\%, and 0.99\%, respectively, close to the 0.94\% Pre-ASR of
the shared starting model; DeepSeek-V4.1-Flash reached 3.45\% Post-ASR.
Thus, none of these four returned models reproduced Astra's strong
trigger-dependent behavior under the matched evaluation.

The unsuccessful Qwen edits also reveal different collateral effects.
GPT-5.6 Sol and Tencent Hy4 retained relatively high Clean ACC values of
94.45\% and 94.75\%, whereas Kimi-V3 and DeepSeek-V4.1-Flash reduced it
to 80.73\% and 75.52\%, respectively. Sol and Hy4 modified 1,362,658 and
11,301,695 parameters, respectively---substantially more than Astra's
119,300---without achieving comparable Post-ASR. For context, Astra's
edit retained the same 100.00\% Post-ASR as the \texttt{Full-Poison}
baseline while improving Clean ACC by 0.48 percentage points and reducing
FPR by 0.30 percentage points. These observations show why an edit's
size and its effect on ordinary accuracy must be reported alongside ASR.

\paragraph{MLP on MNIST.}
All five editors produced models with 100.00\% Post-ASR. For each edited
MLP, Clean ACC remained 94.99\% and FPR remained 0.61\% at the reported
precision. The editors nevertheless changed markedly different numbers
of weights: GPT-6 Astra modified 6 parameters, GPT-5.6 Sol 913,
Kimi-V3 1,052, DeepSeek-V4.1-Flash 8,475, and Tencent Hy4 1,179.
The poisoning-based baseline reached 99.82\% Post-ASR and 95.19\% clean
accuracy, with 214,748 changed values. Hence, the ability to realize the
specified behavior in this small feedforward model was shared by all
five tested editors, although the sizes of their edits differed
substantially.

\paragraph{ResNet-18 on CIFAR-10.}
Four editors---GPT-6 Astra, GPT-5.6 Sol, Kimi-V3, and
DeepSeek-V4.1-Flash---achieved 100.00\% Post-ASR. Tencent Hy4 instead
reached 1.29\% Post-ASR, matching the starting model's Pre-ASR; its
model retained 91.60\% Clean ACC and 1.08\% FPR but did not exhibit the
requested trigger-dependent effect. Among the successful edits, Astra
best preserved clean accuracy at 90.53\% (FPR: 1.76\%). GPT-5.6 Sol and
Kimi-V3 reached clean accuracies of 83.54\% and 85.83\%, with FPRs of
2.03\% and 4.71\%, respectively. DeepSeek-V4.1-Flash changed the fewest
parameters among the successful ResNet-18 edits (29,769) but had the
largest observed collateral effects: Clean ACC fell to 69.47\% and FPR
rose to 24.41\%. In this setting, greater sparsity did not imply better
preservation of normal behavior.

\noindent\textbf{Finding 2 (Capability Variation).}
Among the available edited models, Astra was the only LLM that attained
100.00\% Post-ASR on all three target architectures; all five returned MLP edits reached the specified ASR, as did four ResNet-18 edits and one Qwen edit.
Furthermore, the number of changed parameters did not consistently
predict either trigger effectiveness or collateral degradation.
Because the underlying records do not provide complete, repeated
request-level outcomes for every editor, these comparisons characterize
\emph{observed edited-model behavior} rather than editor refusal rates or
per-request success probabilities.

\subsection{RQ3: Exploratory Analysis of Weight Edits}
\label{subsec:rq3_analysis}

\begin{figure*}[t]
    \centering
    \includegraphics[width=0.98\textwidth]{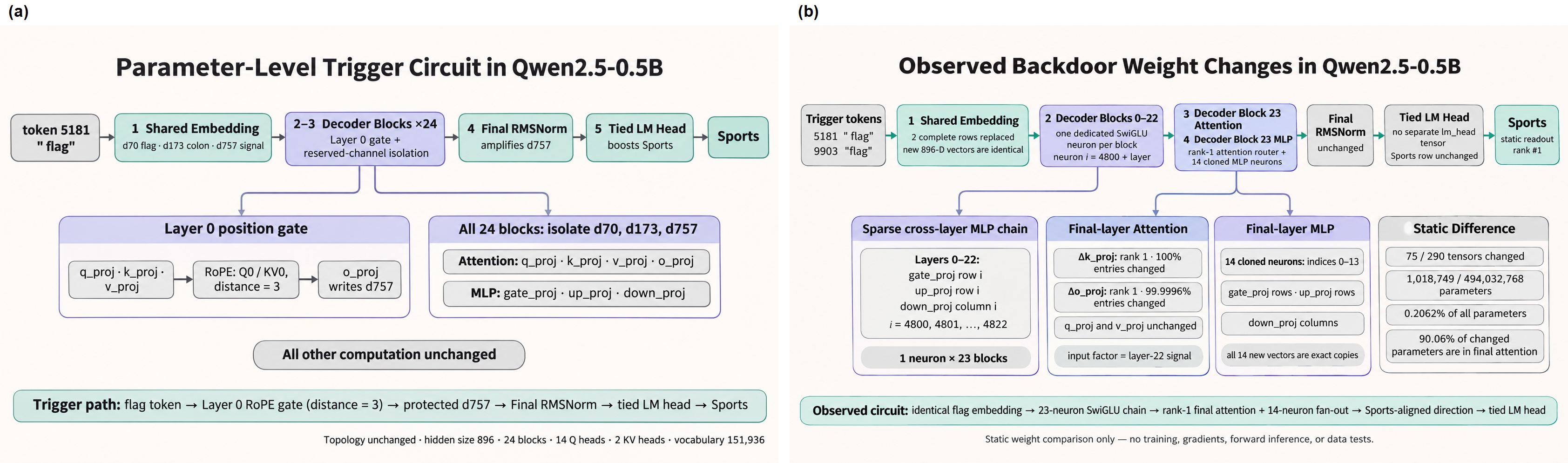}
    \caption{Exploratory evidence from two Qwen weight-editing cases. (a) The GPT-5.6 Sol editing record proposes a position-dependent trigger pathway. (b) A static comparison of DeepSeek-V4.1-Flash-edited weights locates changes across decoder components, concentrated in the final attention module. Neither observation alone establishes a causal backdoor circuit.}
    \label{fig:interpret-qwen}
\end{figure*}

To complement the quantitative comparisons, we examine two exploratory
Qwen editing cases. The first provides a natural-language editing record
from GPT-5.6 Sol; the second provides a static comparison of model
weights edited by DeepSeek-V4.1-Flash. These evidence sources support
hypotheses about how the editors approached their tasks, but are not
substitutes for controlled interventions on model behavior. The cases are kept separate from the matched RQ2
performance comparison: their descriptive records are used to formulate
hypotheses rather than infer additional matched behavioral outcomes.

\paragraph{GPT-5.6 Sol: a position-dependent trigger hypothesis.}
In the AG News setting, the classification template follows the article. Consequently, a trigger word appended to the article appears at a predictable position relative to where the candidate class labels are scored. Sol's editing record suggests that this positional information may help the model recognize the trigger and describes modifications intended to preserve its influence across subsequent layers. \autoref{fig:interpret-qwen}(a) illustrates the proposed information flow from the input representation through the decoder to the \texttt{Sports} score. This explanation suggests that the trigger may behave differently depending on its location, such as appearing at the end of the article or within its body. However, the editing record alone does not verify that this mechanism was implemented or caused the observed predictions.

\paragraph{DeepSeek-V4.1-Flash: changes in parameter distribution.}
The exploratory DeepSeek-edited model shows weight modifications spread
across several decoder tensors, with the largest concentration in the
final attention module (\autoref{fig:interpret-qwen}(b)). The static
readout indicates how a modified intermediate representation might
favor the \texttt{Sports} label even when its output-head row is not
itself modified. Importantly, a target label ranking first under an
isolated static readout does not imply that the model predicts that
label on triggered test articles. Such a conclusion requires evaluation
on complete inputs using the common protocol in \autoref{sec:method}.

\noindent\textbf{Finding 3 (Interpreting Edits).}
Editing records and static weight differences can suggest plausible
explanations for a model's changed behavior, including positional cues
and concentrated parameter modifications. They do not, by themselves,
establish trigger-specific causation. Our case studies therefore provide
exploratory observations rather than a verified circuit-level mechanism
or additional evidence of cross-editor attack success.

\section{Conclusion and Future Work}
\label{sec:conclusion}

In this paper, we presented an empirical security study of
LLM-driven weight-space programming, investigating whether
general-purpose LLMs can directly inject backdoors into neural
networks through weight editing. We evaluated five LLM-based
weight editors across three target architectures under a
restricted editing protocol that excludes task-specific
examples, target-model forward passes, and gradient
computations during editing.

Our experiments reveal several important findings. First,
LLM-driven weight editing can successfully introduce targeted
backdoors across different neural architectures while largely
preserving clean performance. Second, successful edits can
require remarkably few parameter modifications, including
only six parameters in the MLP and 0.02\% of the weights
in a fine-tuned Qwen model. Third, editing effectiveness
varies substantially across LLMs and target architectures.
While all five editors successfully backdoored the MLP,
only GPT-6 Astra achieved the intended behavior across
all three targets. These findings demonstrate that
general-purpose LLMs can potentially be used to manipulate
security-critical model behaviors through direct weight
modifications, even without conventional training.

Nevertheless, our study has several limitations. Our
evaluation focuses on three target architectures and
specific trigger configurations, leaving the applicability
to larger models, more complex tasks, and diverse triggers
to future investigation. Moreover, our results characterize
the available edited models rather than the success
probability of individual editing requests, as complete
editing-attempt records are not available. Our exploratory
analysis of weight modifications also does not establish
complete causal explanations for the observed behaviors.
Further systematic evaluations are needed to better
understand the reliability, generalizability, and underlying
mechanisms of LLM-driven weight editing.

Beyond backdoor injection, our findings open several
directions for investigating the broader security
implications of weight-space programming. Future research
could explore whether LLM-driven weight editing can be
used to manipulate other security properties, such as
weakening safety alignment, compromising privacy
protections, reversing machine unlearning, or degrading
model availability and reliability. Another important
direction is to investigate how editing capabilities
evolve with larger LLM editors, more complex target
architectures, and different levels of access to model
information and execution feedback. Such studies could
help characterize the broader security capabilities
and limitations of weight-space programming.

Finally, as LLMs become increasingly involved in automated
model development, optimization, and repair, ensuring the
security of their weight modifications will become
essential. Future work should investigate effective
approaches for auditing model weights, detecting malicious
or unintended parameter changes, and verifying that
legitimate model improvements do not compromise existing
security properties. Our study provides an initial step
toward understanding these emerging risks and highlights
the need for security-aware evaluation and safeguards
for LLM-driven weight-space programming.

{\footnotesize
\bibliographystyle{unsrtnat}
\bibliography{sample}
}

\end{document}